\documentclass[]{mn2e}

\usepackage{amssymb}
\usepackage{graphicx}
\usepackage{amsmath}
\usepackage{graphics}

\usepackage{color}

\title[diamagnetic silicate polarizers] {On interstellar light polarization by diamagnetic silicate and carbon dust in the infrared}

\author[R. Papoular]{R. Papoular$^{1}$\thanks{E-mail:papoular@wanadoo.fr}\\
$^{1}$Service d'Astrophysique and Service de Chimie Moleculaire,\\
CEA Saclay, 91191 Gif-s-Yvette, France}

\begin{document}

   \maketitle
\label{firstpage}

\begin{abstract}
 The motion of diamagnetic dust particles in interstellar magnetic fields is studied numerically with several different sets of parameters. Two types of behavior are observed, depending on the value of the critical number $R$, which is a function of the grain inertia, the magnetic susceptibility of the material  and of the strength of rotation braking. If $R\leq10$, the grain ends up in a static state and perfectly aligned with the magnetic field, after a few braking times. If not, it goes on precessing and nutating about the field vector for a much longer time. Usual parameters are such that the first situation can hardly be observed. Fortunately, in the second and more likely situation, there remains a persistent partial alignment which is far from negligible, although it decreases as the field decreases and as $R$ increases. The solution of the complete equations of motion of grains in a field helps understanding the details of this behavior. One particular case of an ellipsoidal forsterite silicate grain is studied in detail and shown to polarize light in agreement with astronomical measurements of absolute polarization in the infrared. Phonons are shown to contribute to the progressive flattening of extinction and polarization towards long wavelengths. The measured dielectric properties of forsterite qualitatively fit the Serkowski peak in the visible.

\end{abstract}


\section{Introduction}

Silicates and carbon-rich dust, in some form or another, have long been considered as prime candidate materials for interstellar dust (see, for instance, Spitzer 1978, Whittet 2003 ). It is no wonder, then, that they are also the favorite contenders for polarizers of interstellar light. This paper purports to assess this conjecture in the particular cases of silicates of the forsterite type (Mg2SiO4), graphite and CHONS (complex organic matter) which are all diamagnetic. 

Voshchinnikov  \cite{vos}, like Mathis \cite{mat86} much earlier, made the case for non-iron-bearing silicates, like forsterite. This is mainly based on observational evidence of an anti-correlation, or at least 
no correlation, between polarization and abundance of available iron in the environments that were studied. Nonetheless, and rather unexpectedly, both authors, as well as many others, advocate superparamagnetic iron inclusions in silicate grains as a condition of alignment. However, I have been unable to find any experimental evidence to the effect that the susceptibility of such inclusions have an imaginary component $\chi''$ approaching even the value for paramagnetic substances quoted by Spitzer \cite{spi78}, on the basis of previous work by Jones and Spitzer \cite{jon67}, only considered as ``theoretical insights'' by the authors themselves. To the contrary, van Vleck \cite{van} explains, on p.283, that magnetic friction only occurs through Heisenberg's exchange forces between paramagnetic ions or atoms, which tend to orient the unpaired spins of other atoms. Such forces are negligible in ``magnetically dilute'' media
(p. 294), which is certainly the case of interstellar grains. As for diamagnetism, it is due exclusively to paired electrons , which ``have no resultant spin and so do not give rise to any exchange forces''. Magnetic forces have to be taken into account in the presence of ferromagnetic domains, whose minimum size is of order 100 \AA{\ } and are hardly imaginable in grains.

Only diamagnetic materials are considered below. As a consequence, the Barnett effect (see, for instance, Purcell 1979) cannot be included, since this, like its inverse, the de Haas-Einstein effect, can only be effective in ferromagnetics. For rotation damping, Faraday braking will be invoked here (see Papoular 2017), but will be expressed mathematically so it can be substituted with any other damping torque proportional to the magnetic field.

The form of the equation of motion of a spinning, electrically neutral, grain in a magnetic field in vacuum indicates that there may be two types of fates for the grain according to the value of a critical number, $R$, which depends on the magnetic susceptibility, the moment of inertia and the distribution of electrons over the grain, but not on the field. If $R<10$, the rotation is readily damped and the grain settles into a static steady state in perfect alignment with the field. If $R>10$, the angular velocity is quickly damped until it reaches a critical value, beyond which it decreases very slowly so the grain goes on precessing and nutating about the field for a much longer time. Most realistic values of the parameters of the problem lead to the second alternative (Sec. 2)

Fortunately, in this case, there is still some degree of alignment, as has long been known but attributed to various causes. Here, this is quantified by the average value of cos$^{2}r$ over the run, where $r$ is the angle between the field and the shortest principal axis of the grain. The higher this number, the smaller $r$ is and the better the alignment. This marker is 1/3 for very weak fields, rises up with the field intensity to a maximum, and then quickly decreases. The maximum increases with the rotation braking strength 
and of course, saturates at 1 when the braking allows a final static state to be reached. These statements are illustrated by the detailed analysis of the motion of an a-spherical silicate grain in space. (Sec. 3).

The alignment marker enters the expression of the extinction polarization fraction (degree) as a function of the extinction cross-sections of the grain in its principal orientations relative to the field (Sec. 4). These are computed over the spectral range 0.1-$2\,10^{4}\,\mu$m for forsterite and graphite. The polarization by forsterite compares favorably with observations, while polarization by graphite is more than ten times weaker (Sec. 5).

The effect of collisions with ambient gas atoms is quantified in Sec. 6; it is shown to set upper limits to the polarizing grain size for a given gas density, and to the gas density for a given grain size.

Finally, Sec. 7 presents an interpretation of the observation, by the $Planck$ satellite, of a flattening of the extinction curve and the polarization of the diffuse Galactic medium from far IR to millimeter wavelengths. This is tentatively assigned to the phonon spectrum (coherent vibrations of the whole structure) of whatever material constitutes the grain; the spectrum of a molecule carrying 493 atoms is shown as an example.

\section{Grain motions in a magnetic field in vacuum}

The motion of a spinning particle in vacuum is governed by the vectorial relation (Goldstein 1980)
\begin{equation}
\dot\vec L+\vec\omega\times\vec L=\vec N\, ,
\end{equation}

where $\vec L, \vec \omega, \vec N$ are, respectively, the particle's angular momentum and velocity, and total external torque acting on it.

 \it In the reference frame of the grain, $L_{i}=I_{i}\omega_{i}$ and eq. 1 describes the motion of the field relative to the grain.\rm   Following Purcell \cite{pur79}, let $I_{i}$, where $i=a, b, c$, be the 3 principal moments of inertia of the particle, and $\omega_{i}$, the corresponding rotation speeds; eq. 1 can then be developed into the following set of equations, \it in the Gauss system of units\rm :

\begin{eqnarray}
I_{a}\dot\omega_{a}-\omega_{b}\omega_{c}(I_{b}-I_{c})=N_{a}-I_{a}\frac{\omega_{a}}{\tau_{b,a}}\\
I_{b}\dot\omega_{b}-\omega_{c}\omega_{a}(I_{c}-I_{a})=N_{b}-I_{b}\frac{\omega_{b}}{\tau_{b,b}}\\
I_{c}\dot\omega_{c}-\omega_{a}\omega_{b}(I_{a}-I_{b})=N_{c}-I_{c}\frac{\omega_{c}}{\tau_{b,c}},
\end{eqnarray}

where the second, third and fourth terms from the left of each equation are, respectively the gyroscopic, external and braking torques. Here, $\vec N=\vec M\times\vec B$, is the torque formed by the field and the magnetic moment of the grain. For dia- and para-magnetics, the latter is written $M_{i}=\chi_{i}VH_{i}$, where  $V$ is the overall grain volume. $\chi_{i}$ is the volume susceptibility $\chi_{0}(1-N_{i})$, where $\chi_{0}$ is the bulk susceptibility and $N_{i}$ is the demagnetizing factor along principal axis $i$ (not to be confused with the magnetic torques). Demagnetizing (or depolarizing factors) for the general ellipsoid may be found in  Osborne \cite{osb}; simpler formulas for spheroids are given by Bohren and Huffman \cite{bh}. The sum of the demagnetizing factors along the 3 principal axes is 1. In the Gauss system, magnetic induction $B$, and field, $H$, are expressed in Gauss and Oersted, respectively. Since their numerical values are the same in this system, the term ``field'' will be used for both and given in Gauss. 

In the particular case of rotation braking by the Faraday force (Papoular 2017),

\begin{equation}
\tau_{b,i}=\frac{I_{i}}{s_{i}H<\mid\mathrm{sin}p_{i}\mid>},
\end{equation}

where $p_{i}$ is the angle made by axis i with the field, and $s_{i}$ was shown to be given by

\begin{equation}
s_{i}=\Sigma_{j} (q_{j}d_{j,i}^2)/c\,,
\end{equation}

where $c$ is the velocity of light; the grain is neutral overall, $q_{i}$ is the $j$ th local excess charge (Mulliken's or net charge) and $d_{j,i}$ its distance from the $i$ axis of rotation. For a given material, $s$ scales with the number of atoms, $n_{a}$, in the grain: $s=Sn_{a}$, where $S$ will be designated by ``braking parameter per atom''. Chemical numerical simulations (Papoular 2017) indicated that, for carbon-rich and forsterite (silicate) rich structures, an order of magnitude for $S$ was $10^{-36}$ (escgs). \it If dangling bonds are present on the grain surface or external charges are attached to the grain, they may be included in the sum, and this may considerably reduce the braking time. \rm

Equations 2-4 describe damped rotation as well as oscillator or``pendulum'' motion in the neighborhood of a stable orientation. This is reminiscent of the damped harmonic oscillator albight with notable differences, as the inertia, restoring and damping terms, here, are not constant; moreover, for each of the three dimensions, these  terms depend on the state of the whole system. A numerical solution is therefore the best for each particular case. Nonetheless, the study of several particular cases helps gathering qualitative indications as to the role of the constants of the system in determining the time to reach a steady state closely enough for our purposes, as shown next. 

 \it Stable (unstable) steady state solutions \rm for the motion can be found by nulling the derivatives of the angular velocities. If, moreover, the latter are also null, then, the equations define  states in which the potential energy, $-\vec M.\vec B$, is minimum (maximum). If $M$ is constant (ferro-magnetic material), this occurs when both vectors are parallel. For dia- or para-magnetic materials, this happens when the field is parallel or anti-parallel to that principal axis for which $\mid\chi\mid$ is minimum or maximum, respectively. Examples of these steady \it static \rm states were given in Papoular \cite{pap17}

Another, very popular, kind of steady state has the grain spinning at constant speed. A steady, partially oriented state can be reached provided the grain is spheroidal and the rotation around the symmetry axis, say $c$, is not damped. In the case of electrically neutral dia- and para-magnetics, the second condition  is automatically satisfied as the charges respond instantaneously to the apparent field variations, so $\chi''=0$ (Van Vleck 1932). Under such conditions, there is permanent precession around axis $c$, but no nutation, and the polarization of light by the grain is only partial, depending on the angle made by the symmetry axis with the field. This angle depends upon the grain parameters and its initial conditions, and may span all the range from 0 to 90 deg. Such states will not be considered in further detail here as they require a very exceptional shape.

Whether such steady states can or cannot be reached starting from given initial conditions depends on the grain parameters, for the latter determine the relative weight of each term on the r.h.s. of Eq. 2-4. The gyroscopic term is generally the strongest by several orders of magnitude, but it is conservative and only redistributes rotation energy between the principal directions. The fate of the system is determined by the ratio of the magnetic and damping terms. Multiplying both terms by $\omega_{i}$, this may be thought of as the ratio of energy gained from the field and energy lost to damping. If the former is largely dominant, the system goes on precessing and nutating about the magnetic field. In the opposite case, it swiftly makes its way towards the nearest stable static state in a rather monotonous trajectory. This is made more quantitative by noting that the magnetic term is given by

\begin{equation}
N_{i}=(\chi_{j}-\chi_{k})VH_{j}H_{k},
\end{equation}

so the ratio of magnetic to damping terms is

\begin{equation}
R_{i}=\frac{(\chi_{j}-\chi_{k})VH_{j}H_{k}}{s_{i}H_{i}\omega_{i}},
\end{equation}

neglecting the sine term. 

Starting with a high value, due for instance to a rocket effect, the angular velocity is steadily damped until it reaches  a critical value
$\omega_{i,cr}$, when the grain rotation energy equals its magnetic energy:
 
\begin{equation}
\omega_{i,cr}= (\frac{(\chi_{j}-\chi_{k})VH_{j}H_{k}}{I_{i}})^{1/2}.
\end{equation}

(which is independent of the damping constant and of the initial angular velocity). At that time, $R_{i}$ becomes

\begin{equation}
R_{i}=\frac{((\chi_{j}-\chi_{k})VH_{j}H_{k}I_{i})^{1/2}}{s_{i}H_{i}}\sim \frac{((\chi_{j}-\chi_{k})VI_{i})^{1/2}}{s_{i}},
\end{equation}

to an average factor of order 1. This expression will be designated by the term ``critical number''; note that it slightly differs from that of the $R_{2}$
 defined in Papoular \cite{pap17}. If all 3 $R$'s are smaller than 10, the system proceeds towards one of the 3 stable static states. This is essentially reached at time $t_{f}\sim4\,\tau_{b,i}$. If $R_{i}>10$, it never reaches such a state. In between, it may settle in a pendulum motion about a static state, with maximum excursion $\sim\pi/2$, and maximum angular velocity $\omega_{i,cr}$; the energy gained from the field in one excursion is lost during the next one. One of the $R$'s may happen to be nearly 0, when the grain is very nearly spheroidal, or disc-like as is the coronene molecule. In such a case, it has no choice but to settle in a static state with its axis parallel to the field. This is a particular case of the spinning-precessing heavy top considered briefly above, in which nutation is suppressed and the angle of the symmetry axis with the field decreases to 0 because of rotation braking.

 Note that $R_{i}=\omega_{cr,i}\tau_{b,i}$, and \it is independent of the field as well as of the initial velocity.\rm. In this form, $R$ is seen to be the inverse of the amount by which rotation is damped during one critical period. This helps understanding why its value should not exceed 10 if the grain is to cross the critical rotation frequency  and readily reach a static steady state.

Since $I$ scales like the grain linear dimension to the power 5 , and $V$ and $s$ to the power 3, $R$ scales linearly with the dimension. Thus, larger grains are disadvantaged on both counts: $\tau_{b}$ and $R$; \it perfect alignment with the field  requires smaller grains, weaker susceptibilities and obviously, stronger braking.\rm   In this respect, ice and silicates of the forsterite type are favored as they offer some of the weakest susceptibilities, weaker than that of graphite by a factor 10 (see Dunlop and others 1997).
 
  It should be apparent that this argument is not restricted to the particular case of Faraday braking.

\section{Silicate and carbon grains}
In order to apply these basics to silicates and graphite, we need to know their magnetic susceptibilities and braking properties. The literature on non-iron-bearing minerals (see Dunlop et al. 1997) indicates the following isotropic, volume susceptibilities (in the SI system of units): -80 to -200 for graphite, -13 to -17 for orthoclase, -15 for magnesite, -12 for forsterite (Mg$_{\mathrm{2}}$SiO$_{\mathrm{4}}$, -13 to -17 for quartz, all these values being multiplied by $10^{-6}$; the corresponding values in the Gauss system are 10 times less. Here, using the Gauss system, $-10^{-6}$ will be adopted for silicate, and $-10^{-5}$ for graphite. 

As for the braking parameters, those for Faraday braking of a coronene molecule (36 atoms) were obtained by Papoular \cite{pap17}, using a commercial chemical modeling  package. $s_{i}=1.03, 1.03, 2.07\,10^{-35}$ (escgs). for $i=a,b,c$, respectively (see eq. 6). When the same procedure was applied to a unit cell of forsterite (7 atoms), the following values of the braking parameters were obtained: $s_{i}=3.32, 1, 4.32\,10^{-36}$ (escgs). It will be assumed that a grain of graphite in space is a perfectly disordered agglomerate of coronene molecules, so its braking parameter is isotropic and will be extrapolated from the coronene values; similarly the braking parameter of silicate will be exrapolated from the values for the unit cell.

Consider first a silicate grain. Take $n_{a}$ to be the number of atoms in the grain, and the average volume occupied by an atom to be 2 \AA{\ }$^{3}$. As rough orders of magnitude, take $\Delta\chi=\chi_{j}-\chi_{k}=10^{-7}$ and $S=10^{-36}$ per grain atom. Eq. 10 then gives, in order of magnitude,  $R\sim10\,n_{a}^{1/3}$. It is apparent that, for any realistic grain size, $R$ will be much greater than 10, thus excluding the possibility for any grain to reach a static steady state with any constant orientation relative to the field. A factor 10 or 100 higher $S$ would be welcome. Of course, if any more efficient alternative braking process is available, its braking parameter, $S$, should be substituted in the expression  of $R$, eq.10. 

Meanwhile, one must still inquire about the statistics of the grain direction when a static state is not attainable. Take, for instance, a silicate grain with principal dimensions $a,b,c= 3, 2.5, 0.5\,10^{-7}$ cm, density 1.5, $n_{a}=3308$ atoms in a magnetic field of $3\,10^{-6}$ G.  The demagnetizing factors interpolated from Osborn's curves \cite{osb} are approximately : 0.11, 0.13, 0.76 respectively.

All 3 corresponding values of  $s$ are of order $10^{-32}$, so the 3 braking time constants are, respectively, 2.4, 3.5 and 5.8$\,10^{4}$ s. The critical factor $R$ is equal to 87, 107, and 31 for principal axes $a, b, c$ respectively. The critical angular velocities about axes $a, b, c$ are, respectively, about 9, 7 and 3$\,10^{-3}$ rad/s, and the grain is launched with angular velocity 0.01 rad/s around each axis.

\begin{figure}
\resizebox{\hsize}{!}{\includegraphics{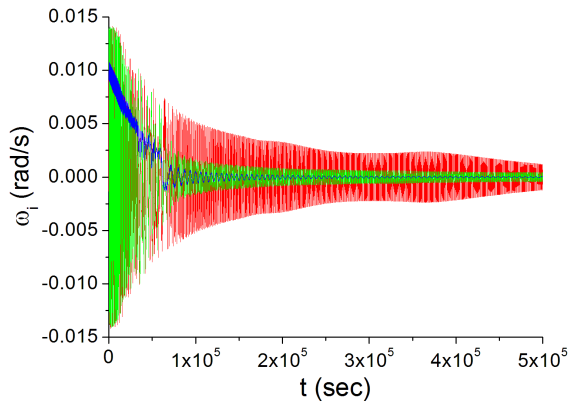}}
\caption[]{An example of behavior, in the grain frame of reference, of the 3 principal angular velocities in time ($a$: red; $b$: olive; $c$: blue), as delivered by the numerical solution of eq. 2-4, with parameters given in the text, such that no static steady state is readily attainable. The maximum values of the three velocities initially decrease steadily due to braking. until about 4$\,10^{4}$ s, when they reach down to the critical velocities. The grain then enters in pendulum mode with  a different amplitude about each principal axis (see Fig. \ref{Fig:Jend} and \ref{Fig:Hsil59b}); their damping constants are much longer than the corresponding braking time, and the longer the higher the value of the critical parameter $R$.}
\label{Fig:w123sil59}
\end{figure}

Figure \ref{Fig:w123sil59} shows the expected behavior: during an initial period, $t\leq 5\,10^{4}$ s, which is about the damping time, the angular velocities decrease until they have been reduced to the critical velocities. This time will be referred to as the critical time, $t_{cr}$. The grain then enters in pendulum mode with  a different amplitude about each principal axis; their damping constants are much longer than the corresponding braking time, $\tau_{b}$, because of the energy periodically fed back by the field.

\begin{figure}
\resizebox{\hsize}{!}{\includegraphics{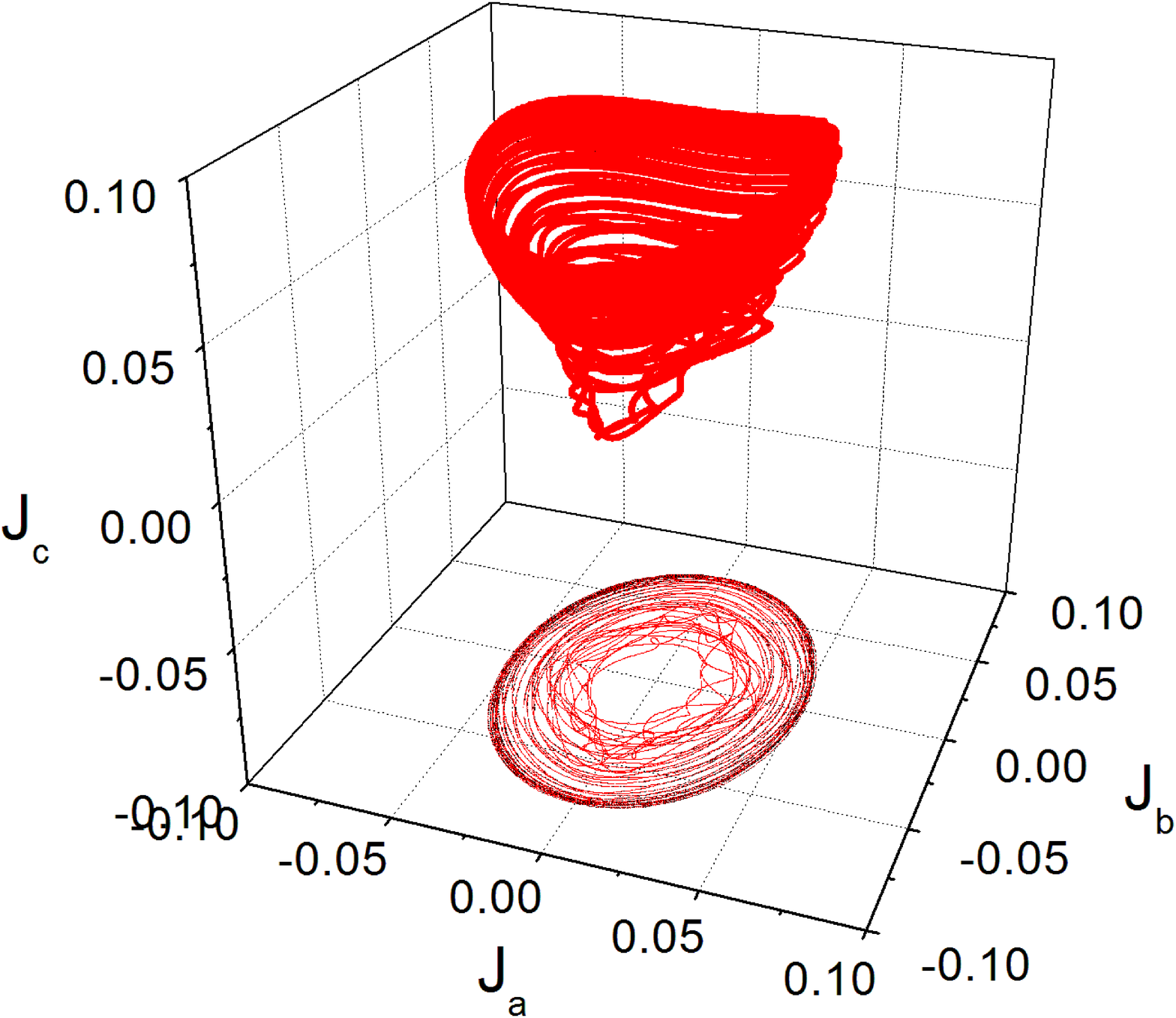}}
\caption[]{The red points describe the trajectory of the tip of the angular momentum vector in the grain's frame of reference during the first part of the run: $t=0$ to $5\,10^{4}$ s, before the initial angular velocities are damped down to the critical velocities (see text); the black points in the $\vec a, \vec b$ plane are its vertical projection. For clarity, only a fraction of the recorded points is included. This illustrates the initial irregular motion, with no preferred orientation.}
\label{Fig:Jbeg}
\end{figure}

\begin{figure}
\resizebox{\hsize}{!}{\includegraphics{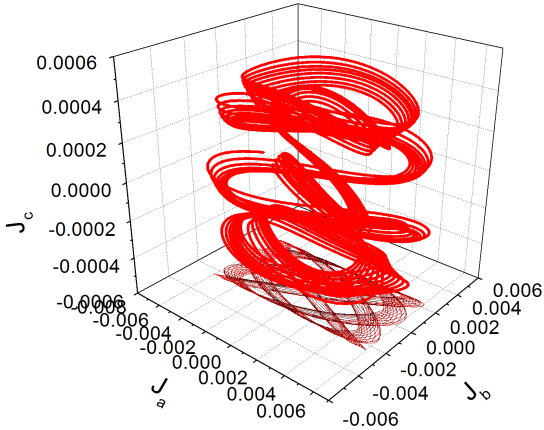}}
\caption[]{
Same as Fig; \ref{Fig:Jbeg}, but for $t>4.5\,10^{5}$ s, after the angular velocities have fallen below the critical velocities. This illustrates the triple pendulum motion, within a restricted range of orientations, during the major part of the run.}
\label{Fig:Jend}
\end{figure}

The initial  internal dynamic evolution is illustrated in Fig.\ref{Fig:Jbeg}, by the trajectory of the tip of the angular momentum $\vec J=J_{a}\vec a +J_{b}\vec b +J_{c}\vec c$ in the grain reference frame, neglecting the off-diagonal terms in the inertia tensor. It is seen to precess about axis $c$ which is special because the grain dimension in this direction is the smallest and the demagnetization factor the largest; this would be the shortest dimension in a prolate grain or oblate grain. As the angular velocities decrease under braking, the precession angle also decreases, until the critical velocities are reached, about $t=5\,10^{4}$ s. Afterwards, the motion is clearly different, as shown by  Fig.\ref{Fig:Jend}, for the last $5\,10^{4}$ s of the run: it is a 3-dimensional pendulum motion, whose amplitudes decrease very slowly.

\begin{figure}
\resizebox{\hsize}{!}{\includegraphics{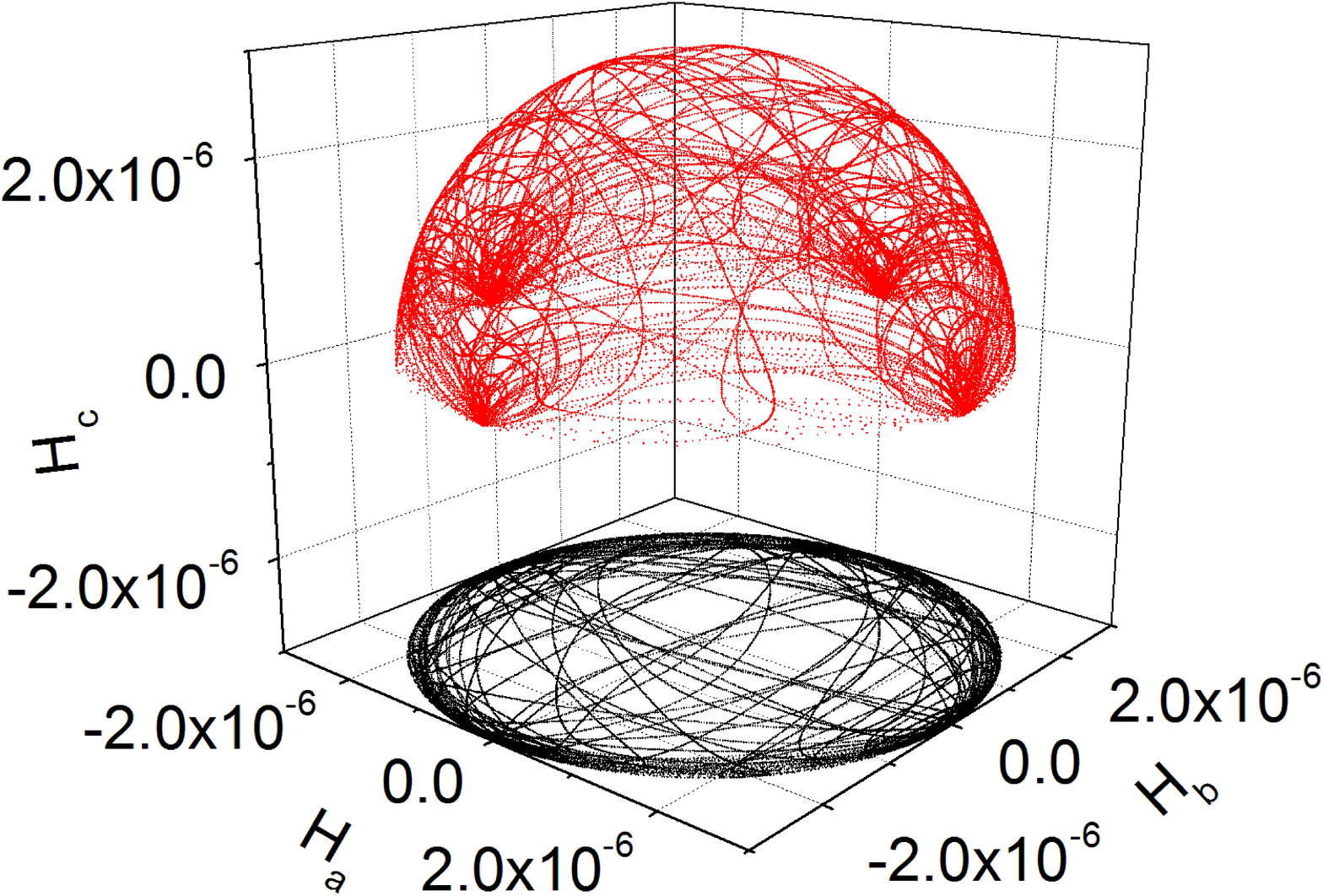}}
\caption[]{The red points describe the trajectory of the tip of the magnetic field vector in the grain's frame of reference during the first part of the run: $t=0$ to $5\,10^{4}$ s, before the initial angular velocities are damped down to the critical velocities (see text); the black points in the $\vec a, \vec b$ plane are its vertical projection. For clarity, only a fraction of the recorded points is included. This illustrates the initial irregular motion, with no preferred orientation.}
\label{Fig:Hsil59a}
\end{figure}

\begin{figure}
\resizebox{\hsize}{!}{\includegraphics{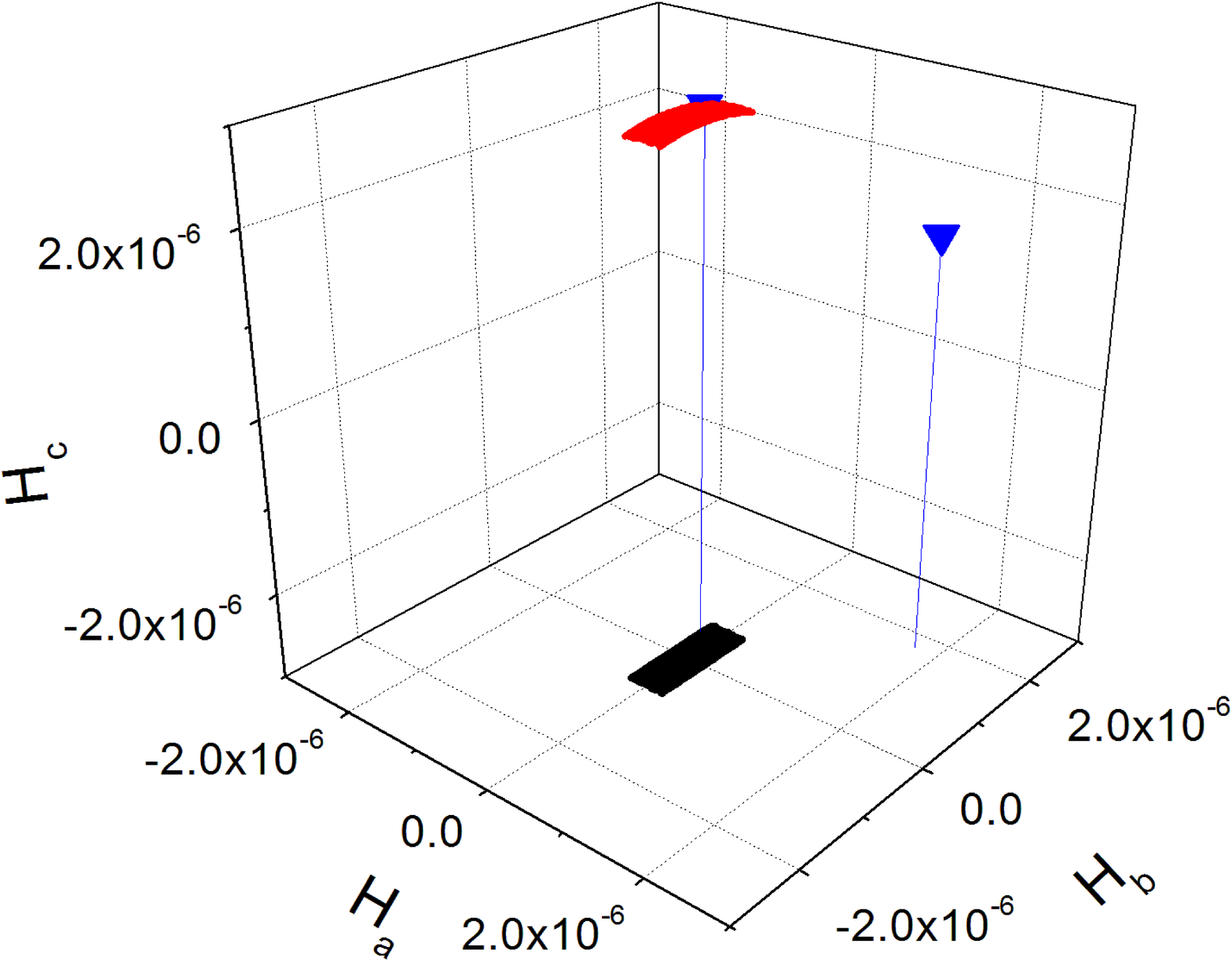}}
\caption[]{Same as Fig; \ref{Fig:Hsil59a}, but for $t>4.5\,10^{5}$ s, after the angular velocities have fallen below the critical velocities. This illustrates the triple pendulum motion, within a restricted range of orientations, during the major part of the run. The large blue squares define the positions of the tip of the field vector at the beginning and end of the run}.
\label{Fig:Hsil59b}
\end{figure}

The evolution of the grain orientation relative to the field (the external dynamics) is illustrated in Fig. \ref{Fig:Hsil59a} and \ref{Fig:Hsil59b}, which, in effect, represent the trajectory of the tip of the field vector in the grain's frame of reference ($\vec a, \vec b, \vec c$). As long as the angular velocities exceed the critical values ($t=0$ to $5\,10^{4}$ s), the grain orientation has no apparent preference relative to the field. Afterwards it is clearly restricted to a small angular range centered on the field direction, where it executes a triple pendulum motion for the major part of the computation run. This preference is justified by the fact that, in this example, axis $c$ bears the strongest demagnetizing factor, and correlatively the highest algebraic susceptibility, which goes with the lowest magnetic potential energy. \it This is independent of the initial angular momentum \rm

The notion of orientation can be quantified with the help of the angles $p, q, r$ made by the field vector with principal axes $a, b, c$. By convention, these angles span the restricted range 0-90 deg, so $r=0$ deg, for instance, means that principal axis $c$ is parallel or anti-parallel to the field, as befits dia-magnetic particles. The ambiguity may be lifted by referring to the 3 field components of Fig. \ref{Fig:Hsil59a} and \ref{Fig:Hsil59b}.

A quantitative and convenient, albeit rough, measure of the degree of orientation  of the grain axis $c$, relative to the field is the \it preference\rm

\begin{equation}
pref_{c}=<\mathrm{cos}^{2}r>,
\end{equation}

where the brackets designate an average over the run time, and similar expressions can be written for angles $p$ and $q$; the squaring of the cosines ensures that  their sum totals 1. If the grain is isotropic or $H=0$, all 3 expressions equal 1/3. If not, they range between 0 and 1. In the example at hand, the average value of  
$pref_{c}$ over the first part of the run is 0.16; over the rest of the run, it jumps to 0.79, when $pref_{a}=0.06$ and $pref_{b}=0.15$. In fact, it steadily increases as the range of pendulum excursions decreases because of rotation braking; thus, during the last $5\,10^{4}$ s, $pref_{c}=0.98$ and the other preferences are reduced accordingly. The dominant parameters among those which determine the preferences are the demagnetization factors, not the braking times, nor the susceptibility, the initial angular velocities or the moments of inertia (just as is the case for steady static states, Papoular 1917).

\begin{figure}
\resizebox{\hsize}{!}{\includegraphics{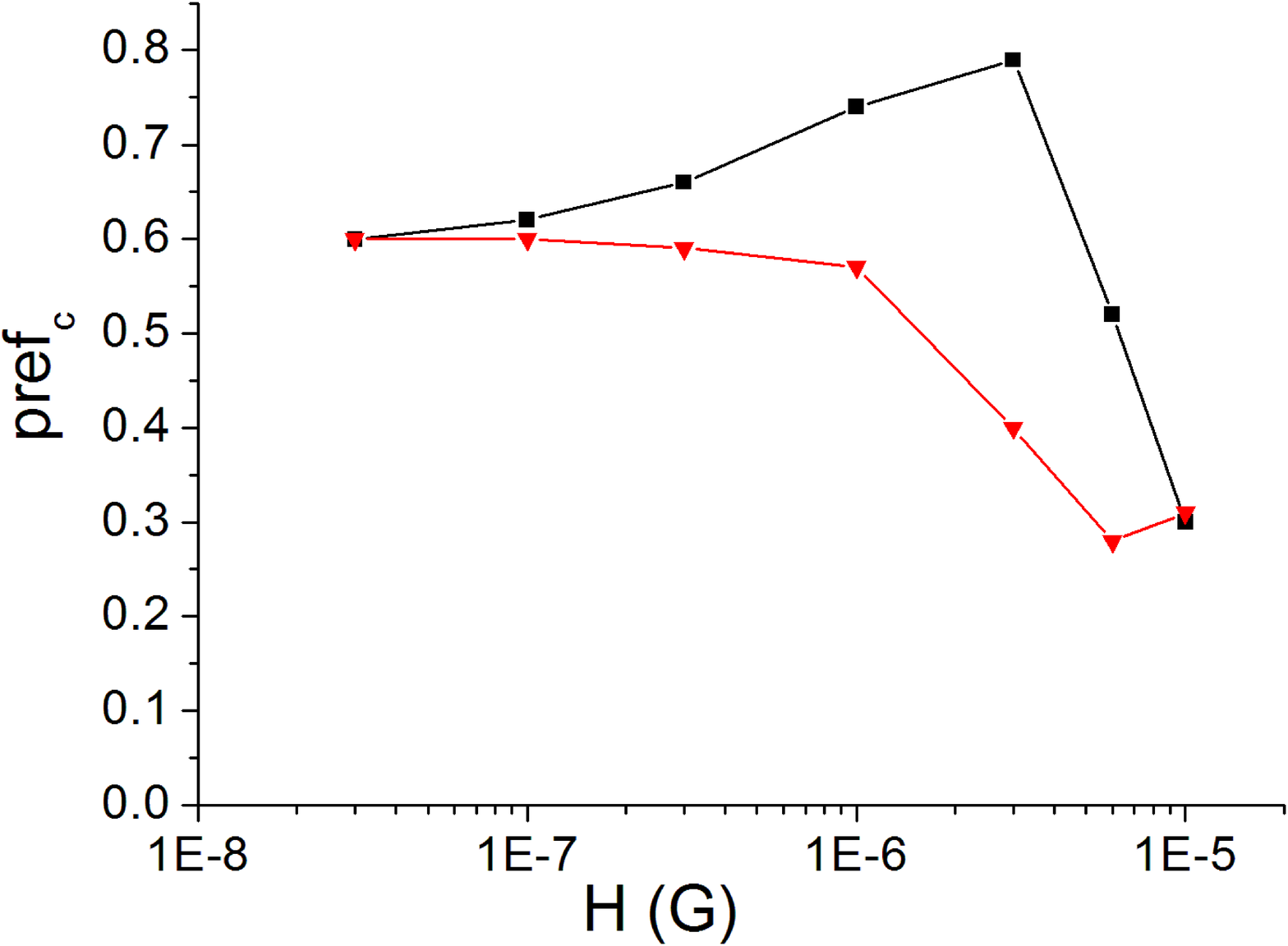}}
\caption[]{The preference $pref_{c}$ for grain alignment with its shorter axis $c$ parallel to the field, as a function of $H$ for 2 values of the braking parameter per atom, $S$: $3\,10^{-36}$ (black squares and line; the line is only for ease of the eye) and $3\,10^{-37}$ (red triangles and line). This preference will enter the reduction factor $RF$ factor in the final expression of the polarization (see Sec. 5). The maximum of the curve increases as $R$ decreases (see Fig. \ref{Fig:Sprefc}. It saturates at 1 when $R<10$. For a given $R$, the preference at low values of $H$ increases with the grain eccentricity. In the limit of strong fields, all orientations tend to be equally probable: preference is 1/3.}
\label{Fig:Hprefc}
\end{figure}

The value of $pref_{c}$ beyond the critical time depends upon the magnetic field as illustrated in Fig. \ref{Fig:Hprefc} (black squares). Starting from $\sim 0.6$ at low field intensities, it rises up to 0.79 at $H=3\,10^{-6}$ G, then quickly drops with increasing field strength. It appears that, in this range, the braking time is so short that the pendulum motions are less and less harmonic so the preference is increasingly blurred. 

\begin{figure}
\resizebox{\hsize}{!}{\includegraphics{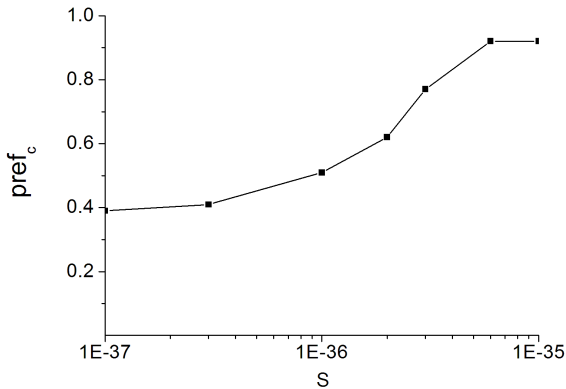}}
\caption[]{The preference $pref_{c}$ for grain alignment with its shorter axis $c$ parallel to the field, as a function of $S$, the braking parameter per atom (black squares), for a constant $H=3\,10^{-6}$ G. The critical number, $R$, decreases as the inverse of $S$.}
\label{Fig:Sprefc}
\end{figure}

 The preference also depends on the braking parameter per atom of the grain, $S$, as illustrated by Fig. \ref{Fig:Sprefc}, at constant $H=3\,10^{-6}$. 

A plausible conjecture is that the behavior can be summarized using the critical number, $R$, which involves the size, the mass, the susceptibility and and the braking parameter per atom: as $R$ decreases, the curve in  Fig. \ref{Fig:Hprefc} rises up; it saturates at 1 when $R<10$ and the grain remains static with the minimum potential energy. 

While the grain chosen above as an example is of a rather oblate type, similar results may be obtained with rather prolate types. In both cases, provided the susceptibility is dia- or para-magnetic and isotropic, the grain orientation will preferentially be such that its shortest principal axis is nearest, if not parallel, to the field vector. Accordingly, the longer axes will be nearly normal to the field, thus ensuring maximum extinction of light polarized normal to the field, which is the configuration favored by observations.    

\section{Absorption cross-sections of nearly spheroidal grains}

In the range of light wavelength and grain size where $2\pi/\lambda<1$, Rayleigh's electrostatic approximation can be applied. This is presented in convenient form by  Bohren and Huffman \cite{bh} (see, also, van de Hulst 1957; Draine and Lee 1984). In their Chapter 5, these authors give the absorption cross-section for a given alignment as

\begin{equation}
C_{abs}=k \mathrm{Im}(\alpha),
\end{equation}

where $k=2\pi/\lambda$, $\lambda$ being the wavelength and $\alpha$ the polarizability of the grain in a given orientation (dimension L$^{3}$). Even for a homogeneous, isotropic material, this depends upon the grain orientation relative to the electric field of the incoming radiation. This dependence is only tractable for ellipsoids and has been derived for the 3 principal orientations, where one of the grain's principal axes is parallel to the wave electric field. For a grain in vacuum, this is

\begin{equation}
\alpha_{i}=V\frac{\epsilon-1}{\epsilon+L_{i}(\epsilon-1)}\,\,;i=a, b, c\,,
\end{equation}
where $V$ is the grain volume, $\epsilon=\epsilon'+\mathrm{i}\epsilon''$ is the complex dielectric function of the material and $L_{i}$, a depolarization factor defined by the ratios of the ellipsoid axes, for alignment of the light's electric field along principal direction $i$ (same as given by Osborn 1945); the 3 factors obey the relation $L_{a}+L_{b}+L_{c}=1$. In the grain example at hand, they are, respectively, 0.11, 0.13 and 0.76.

In order to simplify the polarization computation even further, we shall assume that the grain is nearly spheroidal. For an oblate spheroid with axes $a=b>c$; in the example, the grain dimensions will be approximated by $a=b=2.75\,10^{-7}$ and $c=0.5\,10^{-7}$ cm, so $V=2\,10^{-20}$ cm$^{3}$ and $e=0.97$. The theory then delivers

\begin{equation}
L_{a}=L_{b}=0.12,\,\,L_{c}=0.76\,.
\end{equation}

 Finally, the cross-sections are given by

\begin{equation}
C_{i}=\frac{kV \epsilon''}{(1+L_{i}(\epsilon'-1))^{2}+L_{i}^{2}\epsilon''^{2}}.
\end{equation}

For perfectly aligned grains within a sight line with grain column density $N_{d}$ cm$^{-2}$ to a source of unpolarized light, the observed polarization perpendicular to the magnetic field is
 
\begin{equation}
P= \frac{exp(-\tau_{\perp})-exp(-\tau_{//})}{exp(-\tau_{\perp})+exp(-\tau_{//})}\,,
\end{equation}
where $\tau$ is the wavelength-dependent optical thickness, and $\perp$ and $//$ designate, respectively, light electric vectors perpendicular and parallel to the field.
When the extinction is weak, this can be approximated by

\begin{equation}
P\sim N_{d}(C_{//}-C_{\perp})/2\,,
\end{equation}

where $C_{//}$ and $C_{\perp}$  are the cross-sections of the grain when its symmetry axis is, respectively, parallel and perpendicular to the magnetic field.  In the example of Sec. 3, the approximation is valid beyond 0.1 $\mu$m.

For a grain in permanent rotation and nutation, the polarization is affected by a probability depending upon the time variation of the instantaneous directions of  the 3 principal axes of the grain relative to the reference frame formed by the sight line, the wave electric vector and the magnetic field vector (see Voshchinnikov 2012). \it It will be assumed, here, that this reference frame is orthogonal.\rm By analogy with the Rayleigh reduction factor in common use (see Roberge 2004), we use the preference of Sec. 4 above to define a reduction factor suited to the types of motion described here:

\begin{equation}
RF=\frac{3}{2}(pref_{c}-\frac{1}{3})\,,
\end{equation}

where $pref_{c}$, (Eq. 11), is for axis c to be parallel to the field; $RF=0$ for $pref_{c}=1/3$ and 1 for $pref_{c}=1$.\it The motion of the grain will be taken into account by simply multiplying the expression of $P$ with the reduction factor $RF$\rm,  Although this is a rough approximation, it avoids loosing sight of the underlying physics  

\section{Examples of polarizations}

\begin{figure}
\resizebox{\hsize}{!}{\includegraphics{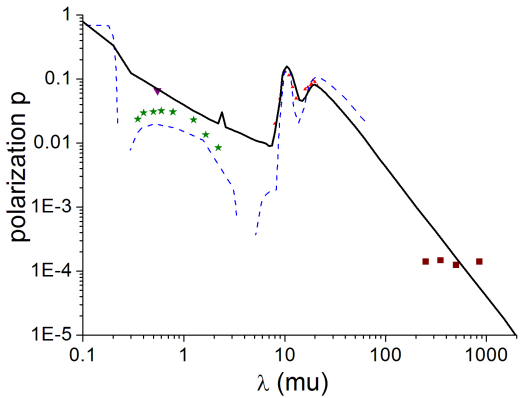}}
\caption[]{The extinction polarization of star light by silicate dust model grains studied in Sec. 3 and 4 (black lines). The solid line was obtained using dielectric functions of silicate grains taken from Draine \cite{dra85} (``astronomical silicate''). The dotted line used the optical indexes of amorphous forsterite measured by Scott and Duley \cite{sco}. \rm The reduction factor for the shortest principal axis $c$ being nearly parallel to the field vector is $RF=0.69$ (see text) and the dust column density was set at $2.3\,10^{15}$ cm$^{-2}$ for astronomical silicate and at $5.75\,10^{15}$ cm$^{-2}$ for amorphous forsterite, so as to fit the measurements of absolute polarization of light from  OMC1 BN by Smith et al. \cite{smi00}, some of which are also represented in the figure (red triangles). The Serkowski peak from HD283701 measured by Whittet et al.\cite{whi92} is also represented here by a few points (olive stars) adapted from their data.A few points (purple squares) in the millimeter  range were deduced  from measurements of diffuse Galactic emission polarization by the Planck Collaboration \cite{adeb} and Ashton et al. \cite{ash}. Similarly, the unique point at 0.55 $\mu$m was deduced from $Planck$ measurements (Ade et al. 2015b).}
\label{Fig:silpol}
\end{figure}

Figure \ref{Fig:silpol} displays the result of polarization calculation for silicate dust as described in Sec. 4, with $pref_{c}=0.79$, which applies to the whole length of the run; the corresponding reduction factor is $RF=0.69$. The solid line was obtained using dielectric functions of silicate grains taken from Draine \cite{dra85} (``astronomical silicate''). The dotted line used data on forsterite from Scott and Duley \cite{sco}. The red, olive and purple symbols are taken, or deduced, from published astronomical observations, and serve as bench marks to the models.

 We discuss first the ``astronomical silicate'' curve in the IR range.\rm  The dust column density was set at $2.3\,10^{15}$ grains.cm$^{-2}$ to fit the measurements of absolute polarization of light from OMC1 BN by Smith et al. \cite{smi00}, some of which are also represented in the figure (red triangles). The fit over the 19-$\mu$m feature is somewhat less satisfactory than that with the 9.7 $\mu$m feature. Since there is 1 Si atom for 7 atoms in the silicate, and each grain carries 3309 atoms, the amount of Si atoms consumed is $N_{Si}=3309*2.3\,10^{15}/7=1.1\,10^{18}$ cm$^{-2}$. 

The BN features are the strongest observed in the sky, which makes them a valuable spectral archetype to compare with. However, the comparison cannot be carried much further as the ``astronomical silicate'' cannot be expected to fit all particular sources. It is interesting, however, to note that Smith et al. measure $\tau(9.7\mu)=1.4$ or 3.2 with a preference for the latter, while the present model gives only $\sim0.64$. This suggests that (partially) oriented grains constitute only 1/5 th of all the silicate grains (and possibly other absorbing materials). Section 6 shows that this may be $partly$ due to collisions with ambient gas atoms. 

The eccentricity of the present model grain was chosen arbitrarily. As an indication, if the 3 dimensions were set at 3, 3 and 2$\,10^{-7}$ cm ($v=7.5\,10^{-20}$ cm$^{3}$), the best fit to OMC BN would require a dust column density $N_{d}=1.5\,10^{15}$ cm$^{-2}$, lower by a factor 0.65, while the quantity of material in a grain increased by a factor 5. This highlights the effect of eccentricity.

The inverted purple triangle at 0.55 $\mu$m is deduced from measurements by the Planck Collaboration (Ade et al. 2015a). They found that the ratio of emission over absorption polarization efficiencies is

\begin{equation}
R_{S/V}=(P_{S}/I_{S})/(p_{V}/\tau_{V})\sim4.3\,,
\end{equation}
where $S$ and $V$ designate the sub-mm and visible bands, respectively, and $P_{S}/I_{S}$ is the emission polarization. They also take $A_{V}=3.1$, so $\tau_{V}=2.8$. The most probable value of sub-millimeter emission polarization of the diffuse IS medium measured by $Planck$ is 10 \% (see Sec. 7). The visible $absorption$ polarization can then be deduced:

\begin{equation}
p_{V}=\frac{p_{S}\tau_{V}}{4.3}=0.07\,,
\end{equation}
which is plotted in the figure, and fits nicely with the solid line. On the other hand, the $emission$ polarization predicted by the ``astronomical silicate model'' for a small optical thickness is 
\begin{equation}
p_{em}\sim \frac{C_{//}-C_{\perp}}{C_{//}+C_{\perp}}\,,
\end{equation}
from which the ratio of sub-millimeter to visible emissions is deduced to be $R_{S/V}'\sim1.5$, smaller than the Planck value, $R_{S/V}$, by a factor 3. This discrepancy may be corrected by reducing $\epsilon'$ of ``astronomical silicate'' by a factor 2 in the spectral range 0.3-3 $\mu$m: this does not affect absorption polarization notably but changes emission polarization enough in this range through the term in $L_{i}\epsilon'$ in the denominator of Eq. 15.

 Changes in the model dielectric functions are also warranted by the less than satisfactory fit to measured absorption polarizations towards isolated individual sources in the visible range. This is illustrated by the Serkowski absorption polarization peak from HD283701 measured by Whittet et al.\cite{whi92}, represented in Fig. \ref{Fig:silpol} by a few points adapted from their data (olive stars). The monotonous descending trend of the model spectrum is obviously at odds with the familiar peak. 

\begin{figure}
\resizebox{\hsize}{!}{\includegraphics{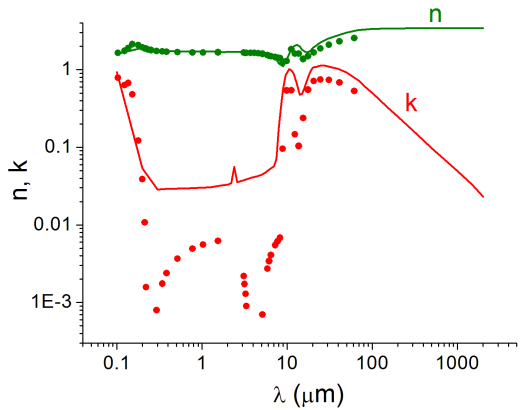}}
\caption[]{Optical indexes:$olive,$ real index $n; red$: imaginary index, $k; lines: $Draine's ``astronomical silicate'' \cite{dr85}, $dots$: adapted from Scott and Duley's \cite{sco} measurements on a forsterite sample. Note the significant drops in Scott and Duley's $k$, at 0.3 and 4 $\mu$m, and the resulting intermediate peak, a factor about 5 below the ``astronomical silicate'' curve, which lowers the corresponding polarization peak by a factor about 4 relative to Draine's data, in Fig. \ref{Fig:silpol}.}
\label{Fig:nk}
\end{figure}

The fit is greatly improved by adopting instead the optical constants $n,k$ measured by Scott and Duley \cite{sco} on a thin film of amorphous forsterite  (Fig. \ref{Fig:nk}). After deducing $\epsilon', \epsilon''$ from these, and substituting them in Eq. 13, I obtained the dotted curve in Fig. \ref{Fig:silpol}. Comparison of Scott and Duley's spectroscopic data with Draine's (Fig. \ref{Fig:nk}) suggests that \it the appearance of a peak in the visible is not due to a  ``resonance'' between $a$ and $\lambda$, but to pronounced drops in $k$ near the two main UV and IR dielectric resonances of forsterite on both sides of the visible range.\rm These drops at very low values of $k$ are washed away in Draine's ``astronomical silicate''. It also becomes clear that minute variations of $k$ in the visible suffice to change the peak polarization value so as to fit observations.

The polarization profile and peak wavelength for a given grain depend on the spectral profiles of $\epsilon'$ and $\epsilon''$ and on the depolarization factor, $L$. $\epsilon''$ is particularly sensitive to addition of impurities, which should therefore be a prime cause of change in this feature. For a population of grains with different shapes, the peak will also depend on the distribution of shapes. Moreover, the dielectric functions themselves may vary according to age or environment. The peak position and, perhaps, even its shape, should therefore be expected to vary from source to source, as is indeed observed (see, for instance, Martin et al. 1999, Martin and Whittet 1999).

 The 4 purple squares in the sub-mm range are also deduced from emission polarization by the Planck collaboration (Ade et al. 2015b). As detailed in Sec. 7, they indicate that the IS polarization levels off in the millimeter range, which requires some correction to the model.\rm

\begin{figure}
\resizebox{\hsize}{!}{\includegraphics{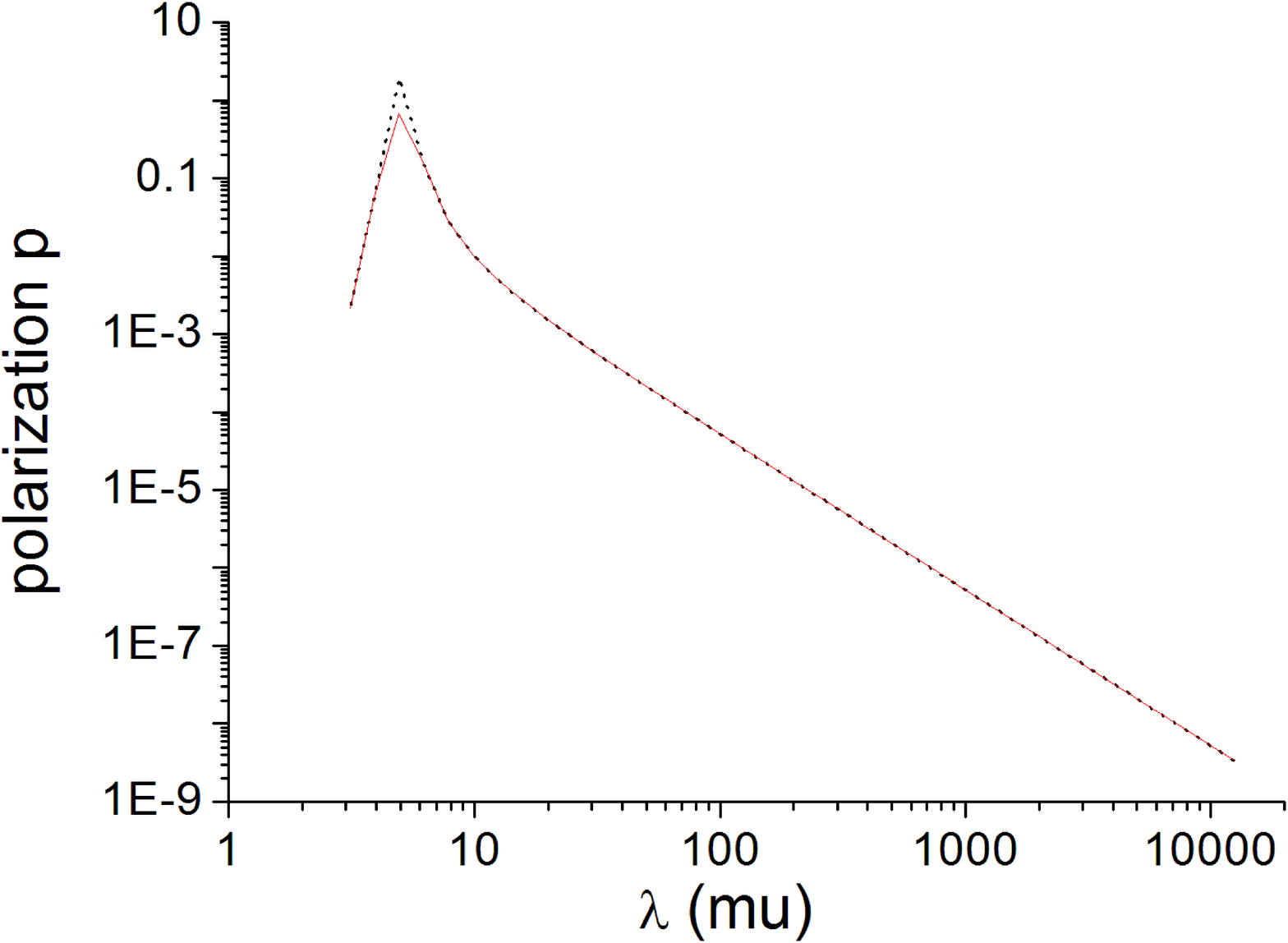}}
\caption[]{The extinction polarization of star light by a grain of the same geometry as the model silicate grain above, only made of HOPG graphite. $red line$: exact expression (16); $black dots$: approximation (17). Same column density of grains, same depolarization factors. The peak near 5 $\mu$m is a surface 
(Frohlich) mode resonance, whose position depends on the electric conductivity of the material and the geometry of the grain; it is absent from the silicate polarization because this material is not a conductor.}
\label{Fig:graphpol}
\end{figure}

Figure \ref{Fig:graphpol} shows the extinction polarization of star light by a grain of the same geometry as the model silicate grain above, but made of HOPG graphite.
The peak near 5 $\mu$m is due to a surface (Frohlich) mode resonance, typical of small grains and whose position depends on the electric conductivity of the material and the geometry of the grain. Because the grain is not spherical, there is a slight spectral shift from one cross-section to the other (see Bohren and Huffman 1983), so the polarization becomes negative below $\lambda=4.5$ (and cannot show up in the logarithmic graph). This phenomenon is absent from the silicate polarization because this material is not a conductor. Such a peak is not observed in the sky; this means that the graphite grains are nearly spherical, or that their absorption cross-section in the mid-IR is much weaker than that of silicates. However, if they are oriented, they might provide an underlying polarization continuum so as to help fit observations, together with silicates  and silicon carbide, for instance (see Gilra 1971)\rm

\section{Effect of collisions with ambient gas atoms}
The analysis of the motion (Sec. 3] showed that whatever alignment is possible for given parameters, is established soon after the critical time, $t_{cr}$, which is the time that is necessary for the the initial angular velocity, $\omega_{0}$ of the grain to be damped down to the critical angular velocity, $\omega_{cr}$ (Sec. 2). For any polarization to be detected at all, the interval between collisions, $t_{coll}$, should be much longer than $t_{cr}$. Now, $\omega_{cr}$ was shown to equal $R/\tau_{b}$, hence,

\begin{equation}
t_{cr}=\tau_{b}\mathrm{ln}(\frac{\omega_{0}}{\omega_{cr}})=2.3\tau_{b}\mathrm{log}(\frac{\omega_{0}\tau_{b}}{R})\,.
\end{equation}

On the other hand,

\begin{equation}
t_{coll}=(f_{inel}n_{H}v_{H}\sigma_{gr})^{-1}\,,
\end{equation}

where  $f_{inel}$ is the fraction of collisions that are inelastic, $n_{H}$ is the density of ambient H atoms, $v_{H}$ is their velocity (corresponding to their temperature, $T_{H}$) and $\sigma_{gr}$ the geometrical area of the grain. Now, to compute these two times, using the example of Sec. 2, where $\tau_{b}\sim5\,10^{4}$ s  and $R$ is of order 100. As for $\omega_{0}$, assume the collision to be inelastic so the grain kinetic energy, $3k_{B}T_{H}/2$ is totally deposited into the grain. Assume, further, that time is available for this energy to be thermalized and equally distributed between the $3n_{a}$ modes of the grain. The 3 rotational degrees of liberty would then have an energy

 \begin{equation}
\frac{1}{2}I\omega_{0}^{2}=\frac{k_{B}T_{H}}{2n_{a}}.
\end{equation}

In the example above, $n_{a}=3309$ and $I\sim10^{-33}$; assuming $T_{H}=20$ K, one finds $\omega_{0}\sim3\,10^{7}$ rad/s, which finally gives $\tau_{cr}\sim10^{6}$ s. Taking $f_{inel}=0.1, n_{H}=20, v=10^{5}$ (cgs) and $\sigma_{gr}=2ab\sim2\,10^{-13}$ cm$^{2}$, one gets

 \begin{equation}
\frac{t_{cr}}{t_{coll}}\sim4\,10^{-3}\,,
\end{equation}

which allows partial polarization to be observed. A fraction of the collisions give rise to surface chemical reaction and formation of a hydrogen molecule, which is readily expelled with a high velocity; in reaction the grain may acquire a high angular velocity, possibly an order of magnitude higher than that estimated above for simple, inelastic collisions. Thanks to the logarithmic function in Eq. 22, the consequent decrease of $t_{cr}$ is not dramatic.

 The ratio in Eq. 25 determines the degree of alignment defined by the reduction factor $RF$, Eq. 18.\rm How does this ratio scale with the grain size $a$, for a given ambient gas density? We note that $t_{cr}$ scales essentially like $a$ and $t_{coll}$ like $a^{-2}$, so the ratio increases like $a^{3}$, which leaves a small range of sizes beyond the 30 \AA{\ } of the Sec. 3, perhaps up to about $10^{-6}-10^{-5}$ cm.

 On the other hand, given a grain size, $t_{cr}/t_{coll}$ decreases as the gas density increases, and so does the alignment degree, in general agreement with observations at all wavelengths\rm. For $a=30$ \AA{\ }, $n_{H}$ could be as high as $\sim$2000 cm$^{-3}$. For a cloud thickness of 10 psec, this would mean a column density of $6\,10^{22}$ cm$^{-2}$ and a visible extinction $A_{V}\sim$30 mag, as upper limits.

\section{Polarization and extinction in the millimeter range}

Equation 17 indicates that, for a single dust material, the spectral trends of both absorption cross-section and polarization should be quite similar. These trends will now be discussed, based on the extensive data delivered by the  $Planck$ satellite. Abergel et al.  \cite{abe14} summarized the emission from dust in the diffuse interstellar medium by defining an absorption cross-section for the spectral range between 100 and 353 GHz (850 to 3000 $\mu$m): 
$(7.1\pm0.6)\times10^{-27}\,\mathrm{cm}^{2}\mathrm{H}^{-1}\times(\nu/353 \mathrm{GHz})^{1.53\pm0.03}$. Comparing this with grey body fits at higher frequencies,
suggests a flattening of the dust spectral energy distribution beyond 850 $\mu$m, which they interpret as possibly due to magnetic dipole emission or to an increasing contribution of carbon dust or a flattening of the emissivity of amorphous silicates at millimeter wavelengths.

Such a flattening seems to be confirmed by measurements of polarization of thermal emission by  Galactic dust by Ade et al. \cite{adeb} at 850 $\mu$m , and Ashton et al. \cite{ash} at 250, 350 and 500 $\mu$m. These authors find an essentially uniform polarization in this spectral range, extending from 0 to $\sim20\%$, with a maximum at 10$\%$ for a hydrogen column density $N_{\mathrm{H}}= 10^{21}$ cm$^{-2}$. The emission polarization can be converted to absorption polarization using 
 
\begin{equation}
p_{abs}(\lambda)=-p_{em}(\lambda)\times\tau(\lambda)\,,
\end{equation}

and $N_{\mathrm{H}}=1.41\,10^{26}\,\tau(353 \mathrm{GHz})$. This gives $p_{abs}(353 \mathrm{GHz})=7\,10^{-7}$ . Assume these relations apply to the silicate dust in OMCI BN, where  $\tau(10 \mu)$=3.2=A$_{V}/18$ (Smith et al. 2000). Then, A$_{V}\sim58$ and N$_{\mathrm{H}}\sim2\,10^{23}$, so 
$p_{abs}(353 \mathrm{GHz})$ rises up to $1.4\,10^{-4}$ and so do the absorption polarizations at 250, 350 and 500 $\mu$m. All 4 are superimposed in Fig. \ref{Fig:silpol}, only to give  a hint as to where the problem is. A hint to one solution is given below.

The descending trend of absorption cross-sections exhibited by the usual dust model spectra is due to the behavior of the dielectric functions with increasing spectral distance from electronic and atomic resonances, This is satisfactorily accounted for by the Lorentz theory of dielectric functions. The latter, however, ignores the phonons, which are coherent vibrations of all the atoms, and are essential in solid state physics for their contribution to the specific heat and thermal and acoustic conductivities of materials (see, for instance, Kittel 1986). Phonons have transverse as well as longitudinal branches, so the former may be excited by light; they may, however, be ignored in the near and mid IR, where their intensity is relatively low, but not in the far IR. The theory of phonons was initially developed for macroscopic crystals, i.e. large, ordered structures. We are rather interested in small, disordered assemblies of atoms. These are best treated like molecules, using chemical modeling codes. 

\begin{figure}
\resizebox{\hsize}{!}{\includegraphics{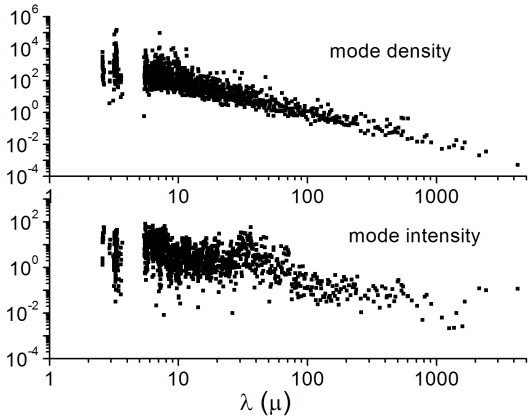}}
\caption[]{$lower \,panel$:The integrated absorption intensities (km/mole) of a CHONS grain including 493 atoms: 254 C, 185 H, 40 O, 5 N and 9 S; $upper \,panel$: the density of modes (per $\mu$m), for a total of 1479 modes. Each point represents a normal vibration of the grain. The absorption cross-section is proportional to the product of mode intensity and density. There is a clear separation between the C-H stretching and the other fingerprint region (6-13 $\mu$m). The 21- and 30-$\mu$m bands can be distinguished. The weaker intensities ($10^{-2}-10^{-1}$ km/mole) are characteristic of phonons, whose domain ranges from the Debye limit near 10 $\mu$ and extends to the mm range. The flattening of the intensity at long wavelengths could be a major cause of the flattening of the IS extinction curve.}
\label{Fig:largemol}
\end{figure}

The IR spectrum of molecules is not due to electronic but to atomic vibrations. The stronger vibrations generally lie preferentially in the near IR, where they are termed ``fingerprints'' and are due to coherent vibrations of small (``functional'') groups of atoms. Underlying these (i.e. having lower intensities), phonons extend from the mid-IR (the Debye cut-off) to wavelengths which scale roughly like the square of the longest dimension of the molecule. They can be distinguished from fingerprints using chemical modeling codes which allow visual representation of atomic motions (see Papoular 2014). With sufficiently large molecules, the vibration spectrum extends beyond the domain of fingerprints, where it carries only phonons. By contrast with fingerprints, their integrated intensity remains essentially constant as the wavelength increases. This is illustrated here in Fig. \ref{Fig:largemol}. Individual phonons are discrete normal modes; their width is narrow at low temperature, only widened by interaction with other modes. However, when a large number of molecules accumulate in the same sightline, the phonons merge into a continuum. A vivid illustration of the phonon continuum may be found in the continuum detected in Spitzer IRS spectra (Smith et al. 2007, Fig. 4): absent in the fingerprint range, it becomes dominant beyond, with a spectral energy distribution depending on dust temperature.

It is argued, here, that the power index of the IS extinction decreases continuously towards longer wavelengths.as a consequence of the presence of both fingerprints and phonons, Papoular \cite{pap14b}  made an extensive study of the phonons of several large molecules: amorphous-carbon type (a-C:H/HAC) and CHONS (see Kwok 2016). The latter are assemblies of some of the more abundant elements in the Universe, having roughly the structure of kerogen, the precursor of coal on Earth. They are also amorphous but contain a sizable fraction of aromatic rings, although they are much less graphitized than HAC. The extinction cross-section of CHONS was found to obey roughly the expression

 \begin{equation}
\sigma_{ext}=5\,10^{-22}\lambda^{-1.4}\,\mathrm{cm}^2\,\mathrm{per H atom}
\end{equation}

in the range $\sim50$ to $500\, \mu$m, which amounts to $2\,10^{-25}$ at $\lambda=250\,\mu$m. In the same range, the cross-section of HAC was found to fall more steeply towards the red, and to be about 2 orders of magnitude weaker. Together with magnesium silicate grains, the presence of 15 $\%$ of cosmic carbon in the form of CHONS was shown to fit the emission spectrum of the Diffuse Galactic InterStellar Medium (DGISM) as measured by the Planck satellite (see Abergel 2014)

 In Papoular \cite{pap14b}, the spectral energy of CHONS was computed as far as 4000 $\mu$m for molecules up to 50 \AA{\ } in length. Their extinction crossection was found to level off progressively beyond a few hundred $\mu$m, and was estimated at roughly $1.7\,10^{-26}$ cm$^{2}$ per Hatom in that range. If only 40 \% of the cosmic carbon is locked in such a material, then it accounts for the total cross-section derived by Abergel et. al. \cite{abe14}, $\sim7\,10^{-27}\,(\nu/353\,\mathrm{GHz})^{1.53}$ cm$^{2}$ per H atom, between 850 and 3000 $\mu$m. As a consequence, it should also account for the flattening of the polarization in the same range (purple squares in Fig. \ref{Fig:silpol}).

It should be stressed here that the existence and general properties of phonon are not specific of the elements constituting the grain nor of the structure being crystalline or disordered. The theory and quantitative treatment can be extended, in particular, to silicate grains.  

 A final note on emission polarization in the FIR and millimeter range. Hildebrandt et al. \cite{hil95} measured emission polarization from molecular clouds, at 100 $\mu$m.
They found it to lie in the range 0 to 10  $\%$, with a median of about 2.2 $\%$, for optical thicknesses from 0 to 0.3. At small thicknesses, their measurements are close to those of $Planck$ for the diffuse medium. This suggests that, at low ambient densities, emission polarization is of order 10 $\%$ in the FIR and millimeter ranges. This can be rationalized as follows. From relation 25 and Eq. 18,
 
\begin{equation}
P_{em}=RF\frac{C_{//}-C_{\perp}}{C_{//}+C_{\perp}}.
\end{equation}

In dilute media, collisions will not hinder alignment, so a population of intermediate sized grains will have a small critical number,$R$, and be able to align perfectly. For these, $RF$ can be taken to be 1. $P_{em}$ then depends only on the ratio 

\begin{equation}
\frac{C_{//}}{C_{\perp}}=\frac{1+L_{\perp}(\epsilon'-1)}{1+L_{//}(\epsilon'-1)},
\end{equation}
which is a function of the real dielectric constant, $\epsilon'$ and the ratio of axes, $c/a$, through the depolarization factors. Taking $\epsilon'=3$ for forsterite, and 0.1 for $P_{em}$, one gets $c/a\sim0.8$. Thus, unlikely high grain eccentricities are not implied by the above observations.\rm

\section{Conclusion}

 This paper advocates no particular mechanism designed to strengthen braking, enhance grain alignment or accelerate rotation. It only seeks to extract indications, and draw conclusions, from the most general possible expression of the exact universal equations of motion in a magnetic field. Any particular torque, such as, for instance, the current popular RAT (radiation torque; see Andersson et al. 2015) can be substituted in the last term of the equations, and the corresponding grain behavior in time be deduced. \rm

1) The numerical solution of the equations of motion of a particle in a magnetic field does not confirm the general validity of the assumption of the constancy of the angular momentum nor the benefit of a distinction between external and internal alignments. However, the consideration of  several different sets of unbiased parameters in these equations reveals different types of motion depending on the combinations of these parameters.

2) Whatever imaginary component exists in the magnetic susceptibility is due to quantum exchange forces between unpaired electrons. It is therefore negligible unless the density of the latter is large, which is the case only with ferro-magnetics. The minimum size of a ferromagnetic domain is of order 100 \AA{\ }; so many such domains are not likely to coexist in the common cosmic dust grain. Dia-magnetics have no unpaired electrons and, therefore, have only a real susceptibility whatever the angular velocity they are given; this is the case of non-iron-bearing silicates and carbon-rich dust.

3) Whatever the initial angular velocity and the type of braking torque, the angular velocity must go through a critical value, when the rotation motion turns into an oscillation about the magnetic field. Beyond that point, the angular velocity  may either a) decrease at the rate defined by the braking torque, and quickly settle in a definite, static,  orientation relative to the field (perfect alignment); or b) go on oscillating for a much longer time (many times the braking time). The difference between the two courses is in the relative amounts of rotation energy lost to braking and energy gained from the field in each cycle. The course that is chosen depends on the critical number $R$, a combination of the susceptibility, the moment of inertia and the ratio of the torque to the field, but not on the field intensity, nor on the initial angular velocity.

4) It appears that, given realistic parameters, perfect alignment is not likely, including precession of the grain axis around the field with a constant angular velocity. Not withstanding, the evolution of the angles made by the field with the 3 principal grain axes reveals a preference for one angle to be smaller than the others (imperfect alignment). It is  the one that has the highest de-magnetization number of the three. An alignment reduction factor, $RF$,  is defined and its dependence on the field and braking is studied numerically. It is shown to increase as $R$ decreases and to equal 1 when $R<10$.

5) For both oblate and prolate grains, perfect or imperfect alignment orientation is such that the shortest principal axis is parallel to the field.  Astronomical observations do not warrant unlikely high grain elongations.\rm

6) Since $R$ scales like the grain size, $a$, and the H collision time like $a^{-2}$, the probability for a grain of size $a$ to be aligned decreases like $a^{-3}$. 
The dispersion of $a$ in the interstellar medium may thus be quantitatively linked to the observed dispersion of polarization measurements as concluded long ago by Goodman et al. \cite{goo} The range of sizes favoring alignment is estimated here to be from a few tens to 1000 \AA{\ }.\rm This helps understanding the polarization data of the $Planck$ satellite, regarding the diffuse IS medium. Also  note that the upper limit of $a$ increases with the rotation braking parameter,$S$ (Sec. 2).\rm

7) The quantitative relevance of these notions to observations is demonstrated in a particular case: OMC BN.

8) The observation, by the $Planck$ satellite in particular, of the progressive flattening of the dust absorption cross-section towards long wavelengths is tentatively attributed to the dominance of phonons of carbon or silicate dust, as opposed to vibration resonances, in this spectral range. Phonons are coherent vibrations carried by any structure, crystalline or amorphous. This can explain some of the absorption and polarization data delivered by the $Planck$ satellite in the millimeter range.

9) Among carbon-rich materials, the so-called CHONS (kerogen-like materials, complex organic molecules, MAONS) are found to have absorption cross-sections high enough to contribute significantly to the observed extinction and polarization in the IR and millimeter ranges. The extinction and polarization of small graphite grains display surface modes which are not observed in the sky; this indicates that they are not prime contributors to IS polarization.

10) The optical constants of forsterite measured by Scott and Duley \cite{sco} are found to fit, at least qualitatively, observed absorption polarizations in the visible and mid-infrared. This is not due to a fortuitous favorable circumstance, but to the presence, on each side of the visible spectral range, of a structural resonance of silicates. The Serkowski peak may thus be explained without invoking the particular behavior of grain sizes of the same order of magnitude as the visible wavelengths. Correlatively, the observed peak shifts may be due to variations of dielectric grain properties with defects or impurities content, rather than to grain size variations.\rm

\section{Acknowledgments}
I am grateful to the anonymous reviewer for justified, helpful and clearly rationalized remarks which helped improve this paper.


\begin{thebibliography}{}
 \bibitem[2014]{abe14}Abergel et al. 2014, A\&A 566, 55
 \bibitem[2015a]{adea} Ade P. et al. 2015, Planck intermediate results XXI, A\&A 576, 106
 \bibitem[2015b]{adeb} Ade P. et al. 2015, Planck intermediate results XIX, A\&A 576, 104 and 107
 \bibitem[2015]{and}Andersson B.-G., Lazarian A. and Vaillencourt J. 2015, ARAA, 53, 501
\bibitem[2017]{ash}Ashton P. et al. 2017, arXiv 1707.02936
\bibitem[1983]{bh}Bohren C. and Huffman D. 1983, Absorption and scattering of light by small particles, John Wiley and Sons, New York \bibitem[1984]{dra84}Draine B. and Lee H. M. 1984, ApJ 285, 89 \bibitem[1985]{dra85}Draine B.1985, ApJ 57, 587
 \bibitem[1997]{dun} Dunlop, David J.; $\ddot{O}$zdemir, $\ddot{O}$zden 1997, Rock magnetism: Fundamentals and Frontiers. Cambridge
Univ. Press. ISBN 0-521-32514-5;  Hunt, Christopher P.; Moskowitz, Bruce P. (1995), "Magnetic properties of rocks and minerals", in
Ahrens, T. J., Rock Physics and Phase Relations: A Handbook of Physical Constants, 3, Washington,DC: American Geophysical Union, pp. 189–204;
O'Reilly, W. (1984). Rock and Mineral Magnetism. Boston, MA: Springer US. ISBN 9781468484687
 \bibitem[1985]{dr85}Draine B. 1985, ApJ Suppl. 57, 587
\bibitem[1967]{jon67}Jones R. and Spitzer L. 1967, ApJ 147, 943
\bibitem[1971]{gil}Gilra D. P. 1971 Nature 229, 237
\bibitem[1980]{gol}Goldstein H. 1980, Classical Mechanics, AW Inc.
\bibitem[1995]{goo}Goodman A., Jones T., Lada E. and Myers P. 1995, ApJ 448, 748
\bibitem[1995]{hil95}Hildebrand H. et al. 1995, A.S.P. Conference Series, vol. 73, p 97, Haas M. et al. eds.
 \bibitem[1986]{kit}Kittel C. 1986, Introduction to solid state physics, J. Wiley $\&$ Sons, N. Y
 \bibitem[2016]{kwo}Kwok S. 2016, Astron. Astrophys. Rev. 24:8
  \bibitem[1999] Martin P. et al. 1999, ApJ 510, 905
  \bibitem[[1999] Martin P. and Whittet D. 1990, ApJ 357, 113
\bibitem[1986]{mat86}Mathis J..1986, ApJ  308, 281
 \bibitem[1945]{osb}Osborn J. 1945, PR 65, 351
  \bibitem[2014]{pap14a}Papoular R. 2014, MNRAS 440, 2396
  \bibitem[2014]{pap14b}Papoular R. 2014, MNRAS 443, 2974
 \bibitem[2017]{pap17}Papoular R. 2017, MNRAS in press
 \bibitem[1979]{pur79}Purcell E. 1979, ApJ. 231, 404 
  \bibitem[1996]{sco}Scott A. and Duley W.W. 1996, ApJ Suppl. 105, 401
\bibitem[2000]{smi00}Smith S. et al. 2000, MNRAS 312, 327
 \bibitem[2007]{smi07}Smith J. et al. 2007, ApJ 656, 770
\bibitem[1978]{spi78}Spitzer L. 1978, Physical processes in the interstellar medium, J. Wiley, N.Y
\bibitem[1957]{vdh}van de Hulst H. C. 1957, Light scattering by small particles, Dover, N.Y.
\bibitem[1932]{van}Van Vleck J. The theory of electric and magnetic susceptibilities 1932, Clarendon Press, Oxford
\bibitem[2012]{vos}Voshchinnikov  N. V. 2012, A\&A 541, A52
\bibitem[1992]{whi92}Whittet D.C.B. Martin P. G. and Hough J. et al. 1992, ApJ 386, 562
\bibitem[2003]{whi03}Whittet D.C.B. 2003, Dust in the Galactic environment, J. Phys. Publ., 2nd ed., Bristol, UK
 \end{thebibliography}
 \end{document}